\documentclass{JHEP3}
\usepackage{epsf,amsmath,amssymb}

\newcommand{\half}{\frac12}

\newcommand{\beq}{\begin{eqnarray}}
\newcommand{\eeq}{\end{eqnarray}}
\newcommand{\nn}{\nonumber}

\newcommand{\new}[1]{{#1}}

\def\al{\alpha}

\def\mn{{\mu\nu}}

\def\be{\begin{equation}}

\def\ee{\end{equation}}
\newcommand{\eref}[1]{(\ref{#1})}

\newcommand{\rem}[1]{}

\def\half{{1\over 2}}
\def\NN{{\cal N}}

\def\nonestar{$\NN=1^*$}

\def\ntwo{$\NN=2$}

\def\nfour{$\NN=4$}

\def\bit{\begin{itemize}}
\def\eit{\end{itemize}}

\def \be {\begin{equation}}
\def \ee {\end{equation}}
\def \bea {\begin{eqnarray}}
\def \eea {\end{eqnarray}}
\def \half{\frac{1}{2}}

\def\THETA{\Delta\chi}
\def\Rs{\sqrt{\lambda}}
\def\Rq{\lambda}

\title{Adjoint Trapping: \\ A New Phenomenon at Strong  't Hooft Coupling}

\author{Sungho Hong$^{ab}$, Sukjin Yoon$^{a}$,
and Matthew J. Strassler$^{a}$\\
$^{a}$Department of Physics and Astronomy\\
P.O Box 351560, University of Washington\\
Seattle, WA 98195\\
\\
$^{b}$Department of Physics and Astronomy\\
University of Pennsylvania\\
Philadelphia, PA 19104-6396}

\abstract{Adding matter of mass $m$, in the fundamental
representation of $SU(N)$, to \nfour\ supersymmetric Yang-Mills
theory, we study ``generalized quarkonium'' containing a (s)quark,
an anti(s)quark, and $J$ massless (or very light) adjoint particles.
At large 't~Hooft coupling $\lambda\gg 1$, the states of spin $\leq
1$ are surprisingly light (Kruczenski {\it et al.}, hep-th/0304032)
and small (hep-th/0312071) with a $J$-independent size of order
$\sqrt{\lambda}/m$.  This ``trapping'' of adjoint matter in a region
small compared with its Compton wavelength and compared to any
confinement scale in the theory is an unfamiliar phenomenon, as it
does not occur at small $\lambda$.  We explore adjoint trapping
further by considering the limit of large $J$.  In particular, for
$J\gg\sqrt{\lambda}\gg 1$, we expect the trapping phenomenon to
become unstable.  Using Wilson loop methods, we show that a sharp
transition, in which the generalized quarkonium states become
unbound (for massless adjoints) occurs at $J \simeq
0.22\sqrt{\lambda}$.  If the adjoint scalars of \nfour\ are massive
and the theory is confining (as, for instance, in \nonestar\
theories) then the transition becomes a cross-over, across which the
size of the states changes rapidly from $\sim\sqrt\lambda/m$ to
something of order the confinement scale $\sim\Lambda^{-1}$.  }

\keywords{con, qcd, ads}

\preprint{hep-th/0410080\\
UW/PT 04-19\\
UPR-1094-T}

\begin{document}

\section{Introduction}

In order to bring AdS/CFT techniques to bear even remotely upon QCD,
the original proposal \cite{Maldacena} must be supplemented by
somehow introducing quarks in the fundamental representation.  This
was first done \cite{Aharony} in the system of D3 branes at an
orientifold 7-plane, which requires the presence of four D7-branes;
the field theory is \ntwo\ $Sp(N)$ with four hypermultiplets in the
fundamental representation and one in the antisymmetric
representation.  However, the technical difficulties, both in the
field theory and in the supergravity, of this system, and prejudices
about what to use it for, obstructed progress for some time.  In
\cite{KK} Karch and Katz cut through these barriers by pointing out
that one could in the AdS context simply add a finite number $N_f$
of \ntwo\ hypermultiplets to $SU(N)$ \nfour\ Yang-Mills, since as
$N\to \infty$ the positive beta function which results is
negligible.  The details of the gauge theory (and the notation we
will use to describe it) are presented in Sec.~2 of
Ref.~\cite{hadsize}; we will not repeat them here.  On the AdS side,
this corresponds to adding a finite number of D7-branes into the
$AdS_5\times S^5$ geometry, and observing that the backreaction of
the 7-branes on the geometry and the dilaton is a subleading effect.

This simple observation then led to a number of interesting
developments.  First, Karch, Katz and Weiner \cite{KKW} showed that
the system exhibits what one might call `Gribov confinement'
\cite{gribov} or simply `strong-field confinement', which involves
confinement of heavy quarks without the presence of flux tubes.  In
\nfour\ Yang-Mills plus massive matter in the fundamental
representation, the absence of flux tubes is evident since the
theory is conformal in the infrared. However, as shown in
\cite{KKW}, if a heavy (s)quark and anti(s)quark of mass $M$ are
pair-created and gradually separated from one another, and if the
theory contains (s)quarks of mass $m<M$, then under some
circumstances a pair of the lighter (s)quarks will be created,
confining the heavy quarks; and this happens even though the theory
is infrared-conformal and does not confine electric flux.

Next, within the same theories, the dynamics of quarkonium states
(and generalized quarkonium containing additional adjoint matter)
were studied in \cite{myers}.  In particular, bound states
containing a squark $Q$ of mass $m$, an antisquark $\tilde Q$ of
mass $m$, and some number $J$ of massless fields $\Phi$ from the
\nfour\ sector were studied, along with their superpartners.  The
spectrum of states for various spins $S$ was studied, with exact
results obtained for $S\leq 1$, which are states described purely by
the eight-dimensional gauge theory living on the worldvolume of the
D7-brane embedded in $AdS_5\times S^5$.  The low-spin states were
seen to be surprisingly deeply bound, with masses of order $m_h
\equiv m/\sqrt{\lambda}$ , where $\lambda \equiv 4\pi g_s N\gg 1$ is
the 't Hooft coupling.

To explore these deeply-bound states further, we recently
\cite{hadsize} studied their form factors with respect to conserved
$U(N_f)$ flavor currents, $SO(4)$ R-symmetry currents, and the
energy-momentum tensor.  To regulate certain computations, we
considered adding masses $m_\Phi$ to the $\Phi$ particles in some
cases.  Among other observations, we discovered that these states all
have size of order $m_h^{-1} = \sqrt{\lambda}m^{-1}$, independent of $J$
and of their radial excitation number.  Moreover, we found that the
states have a size which is not sensitive to the mass $m_\Phi$ of the
$\Phi$ particles (as long as $m_\Phi\ll m_h$.) This is very different
from the weak coupling regime.  A $Q\tilde Q$ state has the physics of
the hydrogen atom (and size $(\lambda m)^{-1}$.)  However, the $Q
\Phi\tilde Q$ state is dynamically more similar to the hydrogen
molecule and has a much larger size; the length scale of the wave
function for $Q$ and $\tilde Q$ is a geometric combination of $m^{-1}$
and $m_\Phi^{-1}$.  In particular, {\it at weak coupling, the size of
the $Q\Phi\tilde Q$ state diverges as $m_\Phi$ goes to
zero,\footnote{This can be seen as follows.  The $Q\Phi\tilde Q$ state
at large $N$ has an attractive Coulomb potential between $\Phi$ and
$Q$, and also between $\Phi$ and $\tilde Q$, but no potential between
$Q$ and $\tilde Q$ (more precisely, there is a repulsive but
$1/N$-suppressed potential.) A straightforward Born-Oppenheimer
calculation shows the $Q$ and $\tilde Q$ move independently and slowly
in a wide and rather flat potential-well induced by the rapidly moving
$\Phi$; it can easily be checked that the overall size of this well
grows as $m_\Phi$ decreases.}  whereas at large $\lambda$ we found
\cite{hadsize} that it remains finite and of order $m_h^{-1}$.}

In short, there is a new phenomenon at large $\lambda$ not previously
observed in gauge theory, in which light particles $\Phi$ are trapped
in a region which is small compared both to their Compton wavelength
$m_\Phi^{-1} $ and to the distance scale at which electric flux is
confined $\Lambda^{-1}$.  This ``trapping'' effect, which we believe
is a new phenomenon and which is related to other unfamiliar stringy
effects at large $\lambda$, is what we will seek to explore further in
this paper.  We will argue that when the number of adjoints in the
generalized quarkonium state becomes parametrically of order
$\sqrt{\lambda}$, the state in the string theory ceases to be a
pointlike gauge boson on the D7-brane; instead it becomes a long
semiclassical string that hangs down below the D7 brane.  This in turn
means that the generalized quarkonium state in the field theory is
becoming large, and the trapping is becoming ineffective.  If the
adjoints are massless, complete untrapping occurs when the number of
adjoints is of order $0.22 \sqrt{\lambda}$.  \new{We will see that
this untrapping transition occurs very rapidly as a function of either
$\lambda$ or $J$; for $J>J_*$ the states are unbound, while for
$J<J_*$ the states of low spin are generally trapped.  If the adjoints
are massive and/or the theory is confining, then the untrapping
transition is interrupted when the size of the generalized quarkonium
state becomes of order the confinement length or the Compton
wavelength of the adjoints, whichever is smaller.\footnote{Our
conclusions regarding the states with large numbers of adjoints differ
significantly from the preliminary suggestions made in \cite{myers}.}
}

Our methods for establishing these claims will be as follows.  First,
we will argue, both on the basis of the spectrum computed in
\cite{myers} and using a BMN-type argument, that the generalized
quarkonium states at large $J$ should become unbound at some $J=J_*$
where $J_*$ is parametrically of order $\sqrt\lambda$.  The BMN
argument confirms, moreover, that these states are not metastable for
$J\sim \sqrt\lambda$, and that a loss of efficient trapping occurs as
$J$ approaches $\sqrt{\lambda}$.  At some point
this allows the generalized quarkonium
states to become much larger than their inverse masses.  We therefore
argue that for $J$ just below $J_*$, the states should be relatively
large in size and have small binding energy.  Such states would
resemble a hydrogen molecule, with fast light adjoints weakly bound to
two heavy (s)quarks.  We therefore expect the motion of the (s)quarks to be
non-relativistic, for $J$ sufficiently close to $J_*$, and therefore a
Born-Oppenheimer-type calculation is appropriate, in which the fast
motion of the adjoints is treated first, allowing for the computation
of an effective potential in which the heavy (s)quarks move, and the slow
motion of the heavy (s)quark states in this potential is treated second.
The computation of the effective potential is simply a Wilson loop
computation, which we carry out in section 3.  \new{This calculation
has its own interest and we explore it in some detail.  As expected, the Wilson
loop shows a gradually decreasing level of trapping, with trapping
entirely lost at $J=J_*=0.22\sqrt\lambda$.  The result
allows us to conclude that the effective
potential in the Born-Oppenheimer calculation for the states
with dynamical (s)quarks has the same transition;}
above a critical value of $J$, the effective potential for the (s)quarks
is simply zero, and no generalized quarkonium bound states can form.
Just below this value of $J$, the effective potential for the (s)quarks
is Coulombic with a small coefficient, so the orbits of the (s)quarks are
hydrogenic, with a computable effective coupling.  This establishes that
our picture for the loss of trapping and the unbinding of the
states is self-consistent.

\section{Trapping Many Adjoints?}

The fact that the trapping effect creates generalized quarkonium
states $\tilde Q \Phi^J Q$ whose size does not
grow with $J$ raises a question as to what happens when $J$ becomes
large compared to $\sqrt\lambda$.  Supergravity cannot describe
these states; instead they are expected to be better
described using a string theory obtained in a BMN limit \cite{BMN}.
Most work on BMN limits has attempted to describe operators in
conformal field theories (or states of field theories on spatial
three-spheres) but a BMN limit describing massive states of nonconformal
confining field theories in four-dimensional Minkowski space was
obtained in \cite{annulon}.  Could a similar BMN limit describe
the massive $S=0,1$ states of the form $Q\Phi^J\tilde Q$ for large $J$?

There is a simple reason to suspect the answer is no.
According to \cite{myers}, the ground state consisting of one
$Q$, one $\tilde Q$ and $J$ {\it massless}
$\Phi$ particles has mass of order $J
m_h = (J/\sqrt{\lambda}) m$.  Clearly when this is greater than
$2m$, the system should not be bound.  However, this reasoning
needs to be checked, especially as large 't Hooft coupling physics has
often held surprises.  In particular, the argument could be correct,
but still there could be a potential barrier which makes these
states metastable and forces them to
decay via tunneling.  To explore this latter
possibility we have computed the BMN limit corresponding to these
states.\footnote{This computation was also discussed briefly
in \cite{myers}.}  As we will see, there is no metastability.

Our setup is that $N$ D3-branes fill the 0123 directions of the
ten-dimensional space, and are located at the origin of the 456789
coordinates.  A D7-brane probe is placed at the position $x^8=
m_Q\alpha'\equiv L$, $x^9=0$, and fills the 01234567 directions. We
can write down the metric in a form that manifestly shows the
embedding of the induced metric on the D7-brane
\begin{equation}\label{metric}
ds^2 = \al' \left[ \frac{\rho^2+ \rho_\perp^2}{\Rs} \eta_\mn dx^\mu
dx^\nu + \frac{\Rs}{\rho^2+ \rho_\perp^2}(d\rho^2 + \rho^2
d\Omega_3^2 + d\rho_\perp^2 + \rho_\perp^2 d\chi^2) \right] \ ,
\end{equation}
where $x^\mu$, $\mu=0,1,2,3$ are the ordinary field-theory Minkowski
space coordinates, and the remaining part of the D7-brane
world-volume is spanned by $x^4,x^5,x^6,x^7$, which we write using
rescaled spherical coordinates $\rho^2 = ( \sum_{i=4}^{7} x_i^2
)/\al'^2 \ $ and $d\Omega_3^2 = d\psi^2 \cos^2\theta + d \theta^2 +
\sin^2 \theta d \Omega_1^2$. The space transverse to the D7-brane is
represented by rescaled polar coordinates $\rho_\perp^2 = (x_8^2 +
x_9^2)/\al'^2 $ and $\chi$. In these coordinates, $\rho$ and
$\rho_\perp$ have mass dimension $+1$, while Minkowski space
coordinates $x^\mu$ have mass dimension $-1$ as usual.

To find a BMN limit for states of definite mass (static states which
are eigenstates of the field-theory Hamiltonian $d/dt$) we should
seek to take a Penrose limit with respect to a null geodesic at a
constant AdS radius, which we should expect to lie near $\rho=0$.
This is because \cite{annulon} hadrons of high charge correspond to
modes which are concentrated close to a nonzero and small AdS
radius.\footnote{A hadron's wavefunction falls off as $r^{-\Delta}$,
where $r$ is the AdS radius and $\Delta$ is the dimension of the
lowest-dimension operator which can create the hadron. A hadron of
large charge $J$ can be created only by an operator of large charge
which, since it contains of order $J$ fields, has $\Delta \sim J$.
Therefore, the wavefunction for a hadron of large charge has a
narrow peak at a small AdS radius.}

\FIGURE[ht]{
\label{fig:falloff}
\epsfxsize=2.0in \epsfbox{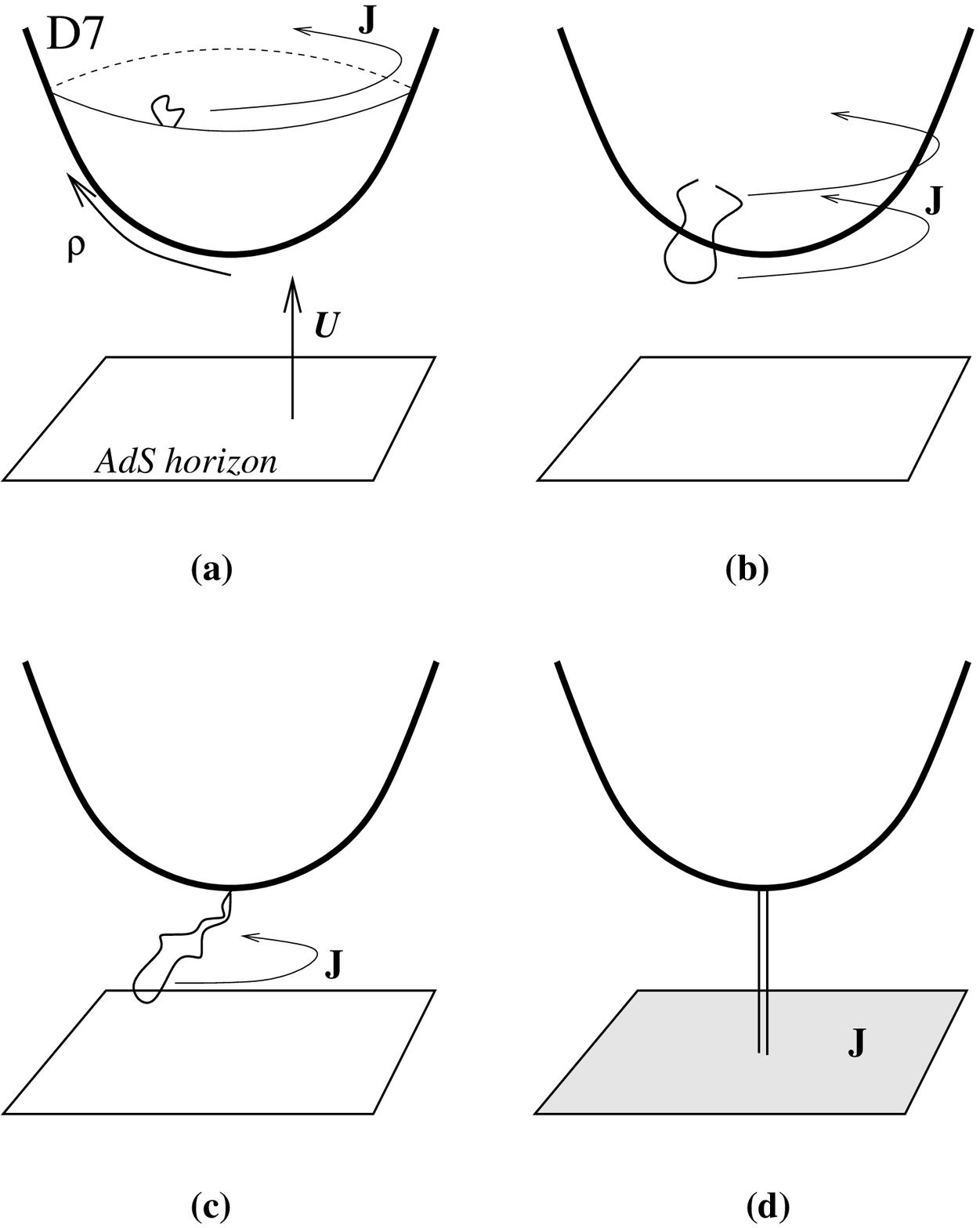}%
\caption{The proposed transition
as the R-charge $J$
(the angular momentum in the horizontal plane) is taken large.
The D7 brane, shown schematically, is
warped by the geometry in the radial direction; gravity
pulls down in the figure, along the coordinate
$U=\sqrt{\rho^2+\rho_\perp^2}$, and the AdS horizon
is shown as a horizontal plane.  (a) A small string with moderate $J$
orbits around
the end of the D7 brane.  (b) \new{As $J\to J_*$}
the string distends and moves off the orbit shown in (a).
(c) The ends of the string descend to the
origin of the D7 brane, while its middle extends below the
D7 brane and carries the angular momentum $J$.
(d)\new{ At $J=J_*$} the string reaches the horizon,
which absorbs the angular momentum,
leaving two long stationary strings --- unbound
(s)quarks.
}
}

We therefore seek
a null geodesic at a constant point in physical space $x^1,x^2,x^3$ and
moving in time and around
an equator of the $S^3$,
{\it e.g.} the curve $\theta =0$.   A particle moving on such a curve
will have both large energy $E$ and large charge $J$.  The effective
Lagrangian for a particle describing this kind of motion is
\[
\mathcal{L}= \al' \left[ - \frac{\rho^2+ \rho_\perp^2}{\Rs}
\dot{t}^2 + \frac{\Rs}{\rho^2+ \rho_\perp^2} (\dot{\rho}^2 +
\dot{\rho_\perp}^2) + \frac{\Rs \rho^2}{\rho^2+ \rho_\perp^2}
\dot{\psi}^2 \right] \ ,
\]
where the $\cdot$ means the derivative with respect to the affine
parameter. The null condition $\mathcal{L} = 0$ gives
\[
\dot{\rho}^2 + \dot{\rho}_\perp^2 + \frac{J^2}{\Rq} \left(\rho +
\frac{\rho_\perp^2}{\rho} \right)^2 = E^2 \ ,
\]
where $E$ and $J$ are the conserved energy and angular momentum
associated with the Killing vectors $\partial/\partial t$ and
$\partial/\partial \psi$ respectively. This is the dynamics of a
particle moving in a potential $V(\rho,\rho_\perp) =\rho \ +
\rho_\perp^2/\rho$. Note that the D7-brane is located at
$\rho_\perp= m_Q$.  If the string itself were fixed to be at
$\rho_\perp=m_Q$, then the minimum of the potential would be at
$\rho=\rho_\perp=m_Q$. However, the string (except for its ends) is
free to move to any value of $\rho_\perp$, and with no constraint on
$\rho_\perp$ the minimum of the potential lies at $\rho_\perp\to 0$.
This shows that there is no stable geodesic with the properties that
we are seeking.  In particular, a particle, or an unconstrained
piece of string, on the above trajectory wants to fall off the
D7-brane toward the horizon of the AdS space.

This result shows the $Q\Phi^J \tilde Q$ states are fully unstable,
not metastable, for $J\gg \sqrt{\lambda}$.  In particular, we are led
to a speculation concerning the nature of that instability. These
states correspond to open strings with both ends on the D7-brane;
while the ends cannot move off the D7-brane, the bulk of the string
worldsheet may be expected, according to the above computation, to
fall off the D7-brane toward the horizon.  This is illustrated in
Fig.~\ref{fig:falloff}.  Once the string becomes an extended object,
and no longer acts as a massless point particle as viewed in ten
dimensions, it will no longer follow the orbit shown in
Fig.~\ref{fig:falloff}a.  (This represents the onset of the loss of
efficient trapping in the gauge theory.) Instead, the string will move
to a lower orbit, with more of the angular momentum being carried by
the lower regions of the string where the energy cost is less due to
the redshift of the metric.  For large enough $J$, the ends of the
string will move almost to the origin of the D7 brane, with the
angular momentum mainly carried by the center of the string, which
continues to orbit around the origin.  As $J$ increases further, the
string will become longer still, and eventually the string will hit
the horizon, at which point the angular momentum will dissolve into
the horizon, as shown in Fig.~\ref{fig:falloff}d.  In this case the
two (s)quarks, represented by two independent straight strings, are
unbound, and the $J$ adjoints are free to move away from the (s)quarks
and from each other.\footnote{To be precise, the colors of the
(s)quark and anti(s)quark are uncorrelated, because they interact at
the planar level only through the adjoints; as the adjoints move away,
the residual force between the (s)quarks is only of order $1/N^2$, and
is repulsive.  Note also that the Minkowski spatial dimensions are of
necessity suppressed in Fig.~\ref{fig:falloff}; the (s)quarks are not
at the same point in space, and the force between them is finite.}

For just slightly smaller $J$, corresponding to the limit of
Fig.~\ref{fig:falloff}c just before it transitions to
Fig.~\ref{fig:falloff}d, the binding of the generalized quarkonium
state is very weak.  In this regime a Born-Oppenheimer calculation
will be valid, in which the ends of the string are held fixed, a
Wilson-loop computation is performed first, and the motion of the ends
of the string is quantized second.  This then motivates us
to carry out the Wilson-loop
calculation, which should show the unbinding transition at $J=J_*$.

\section{A Wilson Loop and the Unbinding of Adjoints}

The Wilson-loop computation in field theory corresponds, as is well
known, to the problem of computing the energy of an appropriate
semiclassical string sitting within a supergravity background.  We
proceed in perfect analogy with the hydrogen molecule calculation of
the effective potential between the two protons induced by the
electrons.  Instead of studying (s)quarks of finite mass $m$ at a
distance of order $1/m$, we will take $m$ to infinity while fixing the
distance $L$ between the heavy squark and antisquark.  Meanwhile we
keep $m_\Phi$ fixed.  If we are in the regime where the $\Phi$
particles are trapped, and $L < m_\Phi^{-1}$, then the $\Phi$
particles should spread out over a region whose size is of order $L$.
If they are not trapped, they should spread out over a region whose
size is of order $m_\Phi^{-1}$.  We will see a transition between
these two behaviors below.  The results of this calculation will give
us the effective potential.  This potential will
remain valid for (s)quarks of finite mass
$m$ if the binding energy $E_0$ of the ground state for finite-mass
quarks in this potential is small compared to $m$.  For $m_\Phi$
sufficiently small, the potential must be of Coulomb type (by
conformal invariance).  A necessary condition for small binding energy
is that the coefficient of the Coulomb term --- the effective $\alpha$
--- be small compared to 1.  \new{This is always true sufficiently
near the transition point $J=J_*$, and this will allow us to compute
$J_*$.  Note however that for any fixed $\lambda$ there need not exist
any bound states well-described by the Born-Oppenheimer method, and
indeed we will see that such states are not generic.}

The easiest way to study this problem initially is to take $m_\Phi$ to
zero.  In this case we need only make appropriate modifications of the
original computation of the Wilson loop by Maldacena \cite{maldaloop}
and by Rey and Yee \cite{reyyee} to determine the coefficient of the
$1/L$ potential between infinitely heavy sources in the fundamental
representation of $SU(N)$ in \nfour\ supersymmetric Yang-Mills.  It
was found that the coefficient was $- 4 \pi ^2 \sqrt{\lambda} /
[\Gamma(1/4)]^4 $ when the sources are antiparticles of one
another. Maldacena also considered the case when the sources couple
differently to the scalar fields of the \nfour\ theory, a possibility
we will return to in just a moment.

To understand the calculation we wish to perform, we need to examine
the theory carefully.  The \nfour\ theory has an $SO(6)$ R-symmetry,
and a sextuplet of scalar fields which it is convenient to organize
into three complex scalars $\Phi_i$.  To preserve the maximal
supersymmetry, a BPS-saturated source in the fundamental
representation of $SU(N)$ must be a massive vector multiplet, which
preserves an $SO(5)$ subgroup of this symmetry.  Alternatively, we
can preserve half the supersymmetry if we add a source which is a
massive \ntwo\ hypermultiplet; this source preserves an $SO(4)$
subgroup of the R-symmetry.  In both cases, the source-antisource
state appropriate to a Wilson loop computation breaks supersymmetry
(in general), but preserves the same $SO(5)$ [or $SO(4)$] subgroup
as the isolated BPS source.  In the computation of
\cite{maldaloop,reyyee}, this source-antisource state is described
as a string whose ends lie at the boundary of AdS a distance $L$
apart, and which are oriented on the $S^5$ so as to preserve the
appropriate $SO(5)$ symmetry [or $SO(4)$.]

We may now ask that the bulk of the string worldsheet be allowed to
rotate around an equator of the $S^5$, so that it picks up a charge
$J$ with respect to an $SO(2)\approx U(1)$ subgroup of the $SO(5)$
[or $SO(4)$.]  This corresponds to the source-antisource state
binding to $J$ complex scalars $\Phi$ which are charged with respect
to this $U(1)$.  (Calculations of this sort of state have also been
undertaken in various papers, especially in
\cite{zarembo,tseytzarembo,annulon} where a very similar method was
needed.)  For the case relevant to the D3-D7 system we have been
considering in this paper, we imagine we introduce an infinitely
massive hypermultiplet $Q, \tilde Q$ which couples to the scalar
field $\Phi_3$ (using a D7-brane at $x^8=m\to\infty, x^9=0$.)  We
may then add $J$ scalars $\Phi_1$ to the state built from a (s)quark
and anti(s)quark by allowing the string to rotate in the
$x^4$--$x^5$ plane.

Since the theory with
$m_\Phi=0$ and $m = \infty$ is conformal \nfour\ Yang-Mills, we know
the potential will be of the form $V(L) = f(\lambda, J)/L$.  Our
simplest goal is to determine $f(\lambda, J)$.  More generally we
wish to determine the shape of the string corresponding to a given
$\lambda$ and $J$.  We will now present this calculation, which
exhibits a trapping transition.

The computation is slightly more complicated than the prototype
\cite{maldaloop,reyyee}, since a third coordinate comes into play.
However, the mathematics largely reduces to one of Maldacena's other
computations, as we will see.  A even more similar computation was
performed by Tseytlin and Zarembo in \cite{tseytzarembo}, and our
techniques follow theirs.  On the surface it would appear that we are
about to repeat their computation, but the details of their solutions
are crucially different.  In particular, we choose different boundary
conditions from \cite{tseytzarembo}.  Our boundary condition
corresponds (as we will see) to a distribution of the global $U(1)$
charge which is regular near the source and antisource; that of
\cite{tseytzarembo} is singular (though integrable.)  The
instability found at weak-coupling in \cite{tseytzarembo} does
not apply to our computation.\footnote{The Euclidean-space calculation
of solution (B) in Sec.~3.2.2 of \cite{tseytzarembo} has mathematical
similarities to our solutions, but arises from a different boundary
condition (again containing a singular charge distribution) and has a
correspondingly different interpretation. To be more precise, the
computations in \cite{tseytzarembo} are done using a boundary
condition that a coordinate called ``$\psi$'' goes to $0$ at the
boundary.  Our coordinate $\theta$ is $\frac{\pi}{2}$ minus this
``$\psi$'', and we instead choose the condition $\theta\to 0$ at the
boundary; this minimizes the global charge density near the source and
antisource and reduces the energy.}

The equations for the shape of the string and its energy are
mathematically equivalent to the computation of \cite{maldaloop} in
which the source and antisource are not each other's antiparticles.
Maldacena introduced two massive vector multiplets which preserved
different $SO(5)$ symmetries, leaving an $SO(4)$.  An angle $\THETA$
enters this computation, describing how badly aligned are the
$SO(5)$s preserved by the source and antisource; more intuitively,
it specifies the angular separation on the $S^5$ of the two ends of
the string as they approach the boundary. For $\THETA=0$ the source
and antisource are antiparticles, while for $\THETA=\pi$ the state
of the source and antisource is BPS-saturated and the binding energy
is zero.  In this case $V(L) = f(\lambda,\THETA)/L$; Maldacena found
an implicit form for the function $f$.  We will see that this
function reappears in the calculation below, but with a very
different interpretation.

\subsection{The Calculation}

We will consider a rectangular Wilson loop of length $L$ and
duration $T$, and we consider the limit $T\rightarrow \infty$ in
order to extract the potential energy between sources at a distance
$L$ from one another. This calculation is dual to a computation of
the energy of a string whose ends lie on the boundary of AdS and are
separated by a distance $L$ in the spatial coordinates of the gauge
theory.

We can use the Polyakov action to describe the semiclassical string
worldsheet
\[
S= - \frac{1}{4\pi \al'} \int d\tau d\sigma
(-\gamma)^{1/2}\gamma^{\al \beta} G_{MN} \partial_\al X^M
\partial_\beta X^N
\]
with appropriate boundary conditions at the end points, which lie at
$\rho_\perp = \infty$.

We are looking for a stationary configuration of the string such
that $\chi=0$. The string should be rotating around an equator of
$S^3$, parameterized by the angle $\psi$; we set $\psi =\omega t$
where $\omega$ is a constant.  We put the (s)quark (one end of the
string) at $X(\rho_\perp = \infty) = L/2$ and the anti(s)quark (the
other end) at $X(\rho_\perp = \infty) = -L/2$, in order that string
configuration be symmetric about $X=0$.  We will find it useful to
employ the coordinates $U$ and $\theta$,
\[ U^2 = \rho^2 + \rho_\perp^2 \quad , \quad
\theta = \arctan\frac{\rho}{\rho_\perp} \ .
\]
$U$ is the usual $AdS_5$ radial coordinate, and $\theta$ is a polar
angle on the five-sphere.\footnote{With the coordinate change, the metric
(\ref{metric}) can be written as
\[
ds^2 = \al' \left[ \frac{U^2}{\Rs} (-dt^2+dx^2)+ \frac{\Rs}{U^2}dU^2
+ \Rs (d\theta^2 + \sin^2\theta\ d\phi^2+ \cos^2\theta\ d\chi^2)
\right] \ ,
\]
which shows that $\theta$ is a polar angle on the full $S^5$.}  We are
interested in solutions in which the source itself is not associated
with any global charge,\footnote{At most, the global charge of the
source should be of order 1, not of order $J$, in order to correspond to
an infinitely massive hypermultiplet.} and so we
expect\footnote{Note the subtleties
addressed in \cite{DGO} do not arise here.} that $\theta\to 0$ and $X(U)$ goes to the solution of
\cite{maldaloop,reyyee} as $U\to\infty$.

Using the conformal gauge $\gamma_{\al \beta} = \eta_{\al \beta}$ ,
the required portion of the string action (setting the string
tension equal to 1) becomes
\[
S = - \frac{T}{4\pi}\int d\sigma \mathcal{L} \equiv TS_0
\]
\begin{equation}\label{L}
\mathcal{L}= \frac{\rho^2+\rho_\perp^2}{\Rs}(\dot{t}^2+X'^2) +
\frac{\Rs}{\rho^2+\rho_\perp^2}(\rho'^2 + \rho_\perp'^2 - \rho^2
\dot{\psi}^2) \ ,
\end{equation}
where $\dot{}\equiv \partial_\tau$ and $'\equiv \partial_\sigma$.
Using the coordinates $U$ and $\theta$, we can separate
$\mathcal{L}$ into two parts which depend only on the $AdS_5$
coordinates and the $S^5$ coordinates, respectively. Fixing
$\tau=t$,
\begin{equation}
\mathcal{L}= \frac{U^2}{\Rs}(1+X'^2) + \frac{\Rs}{U^2}U'^2 + \Rs
(\theta'^2 -\omega^2 \sin^2\theta) \ .
\end{equation}

The equations of motion for the $AdS_5$ coordinates $U$, $X$
decouple from that of the $S^5$ coordinate $\theta$.
\begin{eqnarray}
X'&=& g U_0^2/U^2 \label{eom:X} \\  \nonumber \\
\frac{U'^2}{U^2} + \frac{g^2}{\Rq}\frac{U_0^4}{U^2} -
\frac{1}{\Rq}U^2 &=& \frac{g^2-1}{\Rq}U_0^2 \label{eom:U}
\\ \nonumber \\
\theta'^2 + \omega^2 \sin^2\theta &=& \omega^2 \sin^2\theta_0 \ .
\label{eom:theta}
\end{eqnarray}
Here $g$ is a (dimensionless) constant of integration. At the
midpoint of the string $X=0$, where $\theta'=0$ and $U'=0$, $\theta$
reaches its maximum value $\theta_0$ and $U$ reaches its minimum
value $U_0$. The constraint equation resulting from the conformal
freedom of the worldsheet metric in the Polyakov action,
\[
\frac{U^2}{\Rs}(-1+x'^2) + \frac{\Rs}{U^2}U'^2 + \Rs (\theta'^2 +
\omega^2 \sin^2\theta) =0 \ ,
\]
gives the relation
\begin{equation}
g^2 = 1 - \Omega^2 \sin^2 \theta_0 \quad , \quad \Omega \equiv
\frac{\Rs \omega}{U_0} \ .
\end{equation}

The state we are studying has angular momentum $J$ and energy $E$,
\begin{equation}
J = \frac{\delta S_0}{\delta \omega} = \frac{\omega \Rs}{2\pi} \int
d\sigma \sin^2\theta
\end{equation}
\begin{equation} \label{E}
E = \omega \frac{\delta S_0}{\delta \omega} - S_0 =
\frac{1}{2\pi}\frac{1}{\Rs} \int d\sigma U^2 \ .
\end{equation}
Note that the energy is divergent and must be regulated; we want the
negative potential energy between the source and antisource, so we
must carefully subtract the divergent masses of the sources, as in
\cite{maldaloop,reyyee}.

Remarkably, part of the solution to these equations is of the same form as
one of Maldacena's computations.
In particular, the configuration $X=X(U)$ is the same as that of a
string stretched
between two D3-branes located at the boundary with an {\it angular
separation} on the $S^5$ if we identify $g^2$ in our equations with
$1-l^2$ in equations (4.10)--(4.12) of \cite{maldaloop}.
The string configuration\footnote{Since the string configuration is
symmetric about $X=0$, we describe only the $X>0$ half of the string.} $X =
X(U)$ is obtained from Eq.~(\ref{eom:U})
\begin{equation}\label{eq:X-U}
X = \frac{\Rs g}{U_0 } I(g,U/U_0) \ ,
\end{equation} where
\begin{eqnarray} \label{Ic}
I(g,U/U_0)& = & \int_1^{U/U_0}dy \frac{1}{y^2\sqrt{(y^2-1)(y^2+g^2)}} \\
 & = & \frac{1}{g^2} \left[
E\left(\sin^{-1}(U_0/U),ig\right) -F\left(\sin^{-1}(U_0/U),ig\right)
\right] \ . \nn
\end{eqnarray}
(Here $E$, $F$ and later $K$ are elliptic integrals, with
conventions defined in Appendix C.)
 The string end point is located at $(X,U) =
(L/2,\infty)$, which determines $U_0$ for a given
$g$:
\begin{equation} \label{eq:U0-c}
\frac{L}{2} = \frac{\Rs g}{U_0 }I(g)
\end{equation}
where
\begin{equation}
I(g) \equiv I(g,U/U_0 \rightarrow \infty) =  {1\over g^2}
\left[E(ig) - K(ig)\right] \ .
\end{equation}

The relation between $\theta(U)$ in our computation
and the mathematically-related angular-separation
variable in \cite{maldaloop} is not so direct,
however.
The string configuration $\theta = \theta(U)$ is obtained from
Eqs.~(\ref{eom:theta}) and (\ref{eq:X-U}):
\begin{equation} \label{eq:theta-U}
\int_0^\theta
\frac{d\theta}{\sqrt{1-\frac{\sin^2\theta}{\sin^2\theta_0}}} =
\sqrt{1 - g^2}\int_{U/U_0}^{\infty}
\frac{dy}{\sqrt{(y^2-1)(y^2+g^2)}} \ .
\end{equation}
Here we used $\theta(U \rightarrow \infty) \rightarrow 0$,
consistent with the boundary condition.\footnote{Curiously, the
equations also allow for a solution with $\theta\to \theta_0$ as
$U\to\infty$, one which has the same relation between $g$ and
$\theta_0$ and which has the same $J$ and $U(X)$. However, the
energy of such a state is infinitely larger than the one of interest
to us, reflecting the fact that a source with fixed nonzero $\theta$
is a much longer string than one with $\theta=0$ fixed.} While the
right-hand side of this equation appears verbatim in equation (4.10)
of \cite{maldaloop}, the left-hand side is significantly different
in all respects.

Equation (\ref{eq:theta-U}) may also be written as
\begin{equation}
\sin\theta_0 \
F\Big(\sin^{-1}(\sin\theta/\sin\theta_0),\sin\theta_0\Big) =
\sqrt{1-g^2}\ F\Big(\sin^{-1}(U_0/U),ig\Big) \ . \nn
\end{equation}
The condition that $\theta=\theta_0$ at $U=U_0$ gives the relation
which determines $g$ for a given $\theta_0$:
\begin{equation} \label{eq:theta-c}
\sin\theta_0\ K(\sin\theta_0)= \sqrt{1-g^2} K(ig) \ .
\end{equation}

Using the equations of motion,
the angular momentum $J$ can be
written as
\begin{eqnarray}
J &=& \frac{\Rs}{\pi}\int_0^{\theta_0}d\theta \frac{\sin^2\theta}{\sqrt{\sin^2\theta_0 -\sin^2\theta}} \nn \\
&=& \frac{\Rs}{\pi}\Big[K(\sin\theta_0) - E(\sin\theta_0)\Big] \ .
\label{eq:J}
\end{eqnarray}
The energy in Eq.~(\ref{E}) should be regularized by subtracting the
masses of $Q$ and $\tilde Q$ because it includes infinite $Q$ and
$\tilde Q$ masses.
\begin{eqnarray}
EL & = & \frac{U_0}{\pi} \left[ \int_1^\infty dy \left(
\frac{y^2}{\sqrt{(y^2-1)(y^2+g^2)}} -1 \right) -1 \right] \nn \\
& = & - \frac{2\Rs g^3}{\pi} [I(g)]^2 \label{eq:E}
\end{eqnarray}
which again matches equation (4.13) of \cite{maldaloop}, but
with quite a different interpretation, as we will now see.

Our solution is now complete: we can make one-to-one correspondence
between the parameters in our equations with the physical
quantities, $J$ and $E$. For a given $J$, there is a corresponding
$\theta_0$ from the relation Eq.~(\ref{eq:J}) and $g$ is determined
from Eq.~(\ref{eq:theta-c}). Accordingly, $E$ is obtained from
Eq.~(\ref{eq:E}). Finally, $U_0$ is determined from the relation
Eq.~(\ref{eq:U0-c}), and $U(X)$ can be implicitly found from
Eq.~(\ref{eq:X-U}).

However, it can be seen that the range of $J$ is restricted, and
cannot be arbitrarily large.  This is the sign of the instability
which we have been seeking. The left side of Eq.~(\ref{eq:theta-c})
is an increasing function of $\theta_0$, while the right side is a
decreasing function of $g$. The right-hand side of
Eq.~(\ref{eq:theta-c}) has an upper-bound of $\pi /2$, which occurs
when $g \rightarrow 0$, while the left-hand side diverges
logarithmically as $\theta_0 \rightarrow \pi/2$.  Therefore, there
exists an upper bound for $\theta_0$, which we will call $\theta_*$,
with
\begin{equation}\label{eq:theta*}
\sin\theta_*\ K(\sin\theta_*) = \frac{\pi}{2}
\end{equation}
and correspondingly
one for $J$, which we will call $J_*$:
\begin{equation*}
J_* = \frac{\Rs}{\pi}\Big[K(\sin\theta_*) - E(\sin\theta_*)\Big] \ .
\label{eq:J*}
\end{equation*}
 Numerically, $\sin\theta_*
\approx 0.793$ and $J_* \approx 0.22 \sqrt{\lambda}$~.

To know what does happen when $J \rightarrow J_*$, we need to analyze
the behavior of the string\footnote{In this limit, $\omega \rightarrow
U_0/(\Rs \sin\theta_*) \sim g \rightarrow 0$. Note $\omega$ is not a
physical quantity; it is $J$ that is physical, and larger $J$ does not
necessarily means larger $\omega$.} as $g \rightarrow 0$.  Since $I(g)
\rightarrow \pi/4$ as $g \rightarrow 0$ ($J \rightarrow J_*$),
\be\label{scaling}
\quad U_0 \sim g \rightarrow 0 \quad ,\quad EL \sim -g^3 \rightarrow
0 \ .
\ee
This shows that $U_0$ touches the horizon and the interaction energy
vanishes as $J \rightarrow J_*$. \new{That is, for fixed $L$,
\emph{the massless $\Phi$
particles completely unbind from the infinitely
heavy (s)quark and anti(s)quark for $J>J_*$.}
As we will discuss further below, this implies  that generalized
quarkonium states also become unbound (for massless adjoints)
for $J$ greater than $J_*$.}

\FIGURE[t]{
\epsfxsize=3.5in \epsfbox{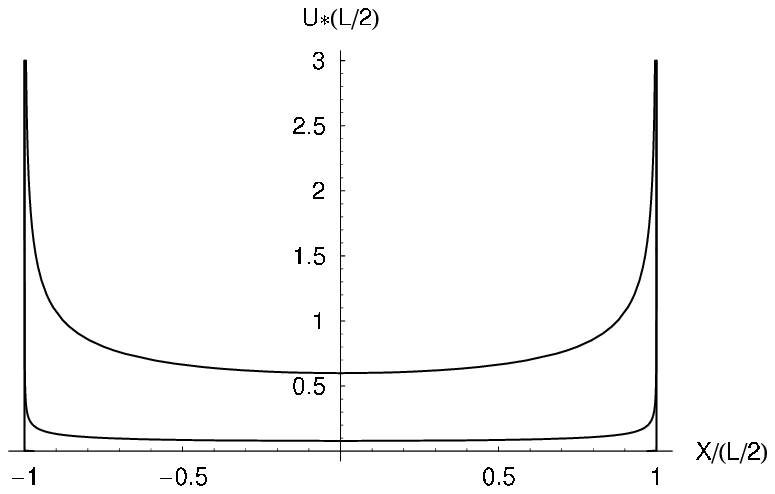}%
\caption{String configurations for $g=1$(dotted), $0.1$(thin solid),
and $0$(thick solid) (equivalently, $J/J_* = 0$, $0.9867$, and $1$,
respectively) in the $X$-$U$ plane. Note the minimum value of $U$ is
a decreasing function of $J$.}%
\label{fig:XU}%
}

\FIGURE[t]{
\epsfxsize=3.7in \epsfbox{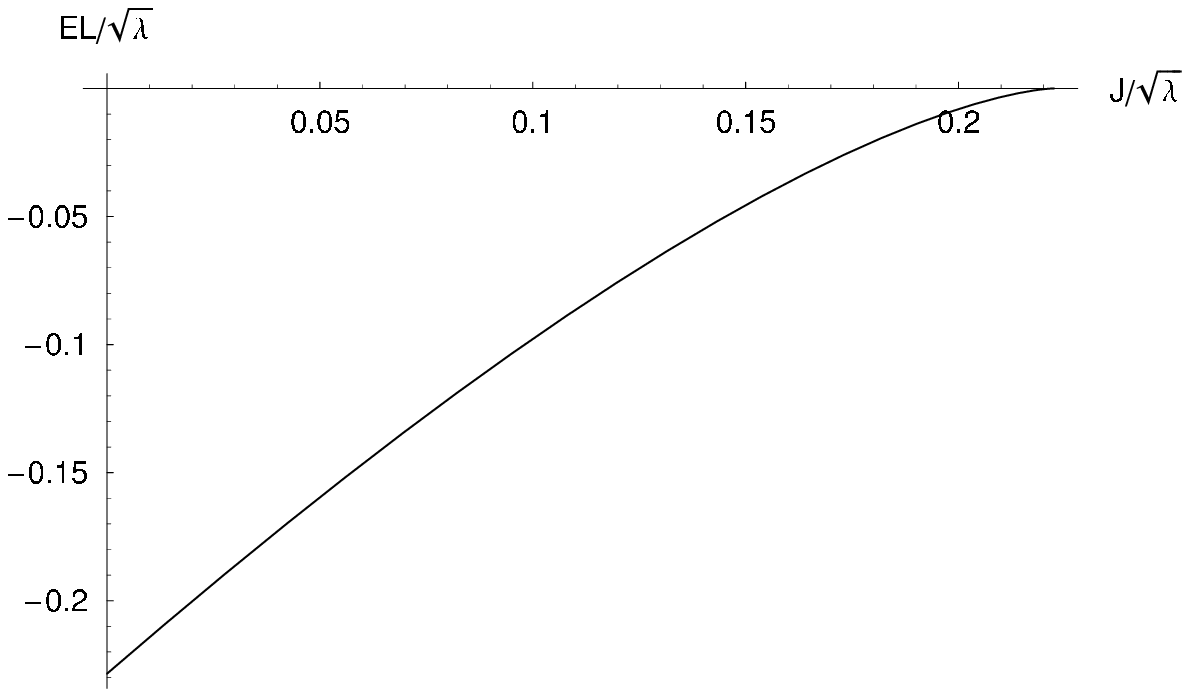}%
\caption{$E$ versus $J$}
\label{fig:EJ}%
}

The relation between $E$ and $J$ \new{(for fixed $L$)} is plotted numerically in Figure
\ref{fig:EJ} and is obtained in two limits ($J \rightarrow 0$ and
$J\rightarrow J_*$) in Appendix A; the result is
\begin{eqnarray}
EL  &\approx&
-\frac{4\pi^2\Rs}{\left[\Gamma\left(\frac{1}{4}\right)\right]^4}
\left(1-2\pi\frac{J}{\sqrt{\lambda}}+\dots\right)
\quad , \quad J \rightarrow 0 \\
EL  &\approx&  - \pi \left(\frac{2}{3}\right)^{3/2}\sqrt{\lambda}
\left[ \left(\frac{J_* -J}{\sqrt{\lambda}\sin\theta_* }\right)^{3/2}
+\dots \right] \quad , \quad J\rightarrow J_* \ . \label{ELJstar}
\end{eqnarray}

Note $|E|$ decreases as $J$ increases,
as expected; the presence of the scalars in the state makes the
binding energy smaller in magnitude.  Interestingly, the relation
between $E$ and $J$ is nonanalytic as $J\to J_*$, indicating a sharp
transition.

It is interesting to ask where the scalar fields actually lie in the
space between the source and antisource.  We do not know the precise
answer to this question, but some significant insight is obtained by
examining the density of $J$ as a function of $x=X/[L/2]$, or as a
function of $y=U/U_0$. For a given $g$, the density of $J$ is given
by
\begin{equation} \label{eq:Jy}
\rho(y) \propto \frac{\sin^2\theta(y)}{\sqrt{(y^2-1)(y^2+g^2)}}
\quad , \quad \rho(x) \propto y^2 \sin^2\theta(y) \ ,
\end{equation}
where $\theta(y)$ is given in Eq.~(\ref{eq:theta-U}).  This is
illustrated in Figure~\ref{fig:Jdensity}. As can be seen, as $J\to
J_*$ the scalars cluster at $U\sim U_0\to 0$; using the usual
inverse relationship between the $U$ coordinate and the size of
distributions in the field theory \cite{Maldacena,peetpolch}, this
strongly suggests that the scalars spread out, and thus the size of
the state is diverging, as $J\to J_*$.\footnote{In the
Minkowski-space solution of \cite{tseytzarembo}, the factors of
$\sin^2\theta(y)$ are replaced with $1$.  This leads to more
singular distributions of charge, since $\sin\theta\sim 1/y$ at
large $y$.}

\FIGURE[h]{
$\begin{array}{c @{\hspace{0.25in}} c}%
\epsfysize=1.8in \epsfbox{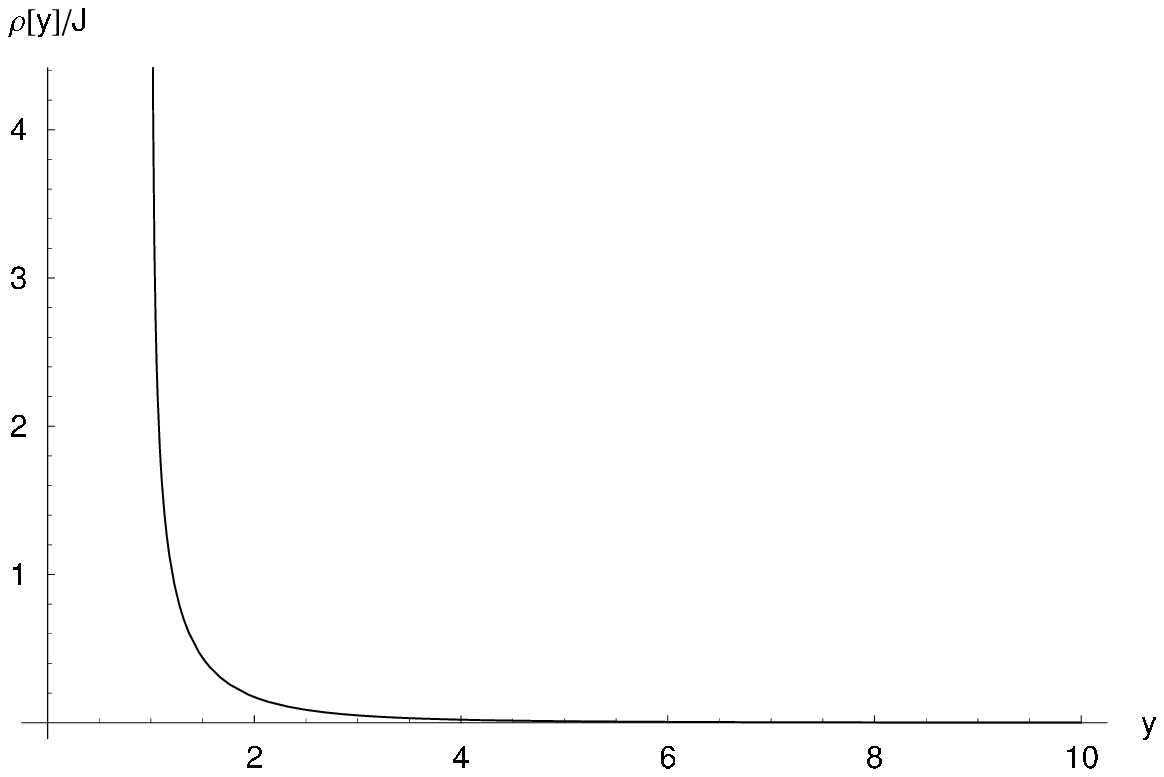} & %
\epsfysize=1.7in \epsfbox[93 -3 428 198]{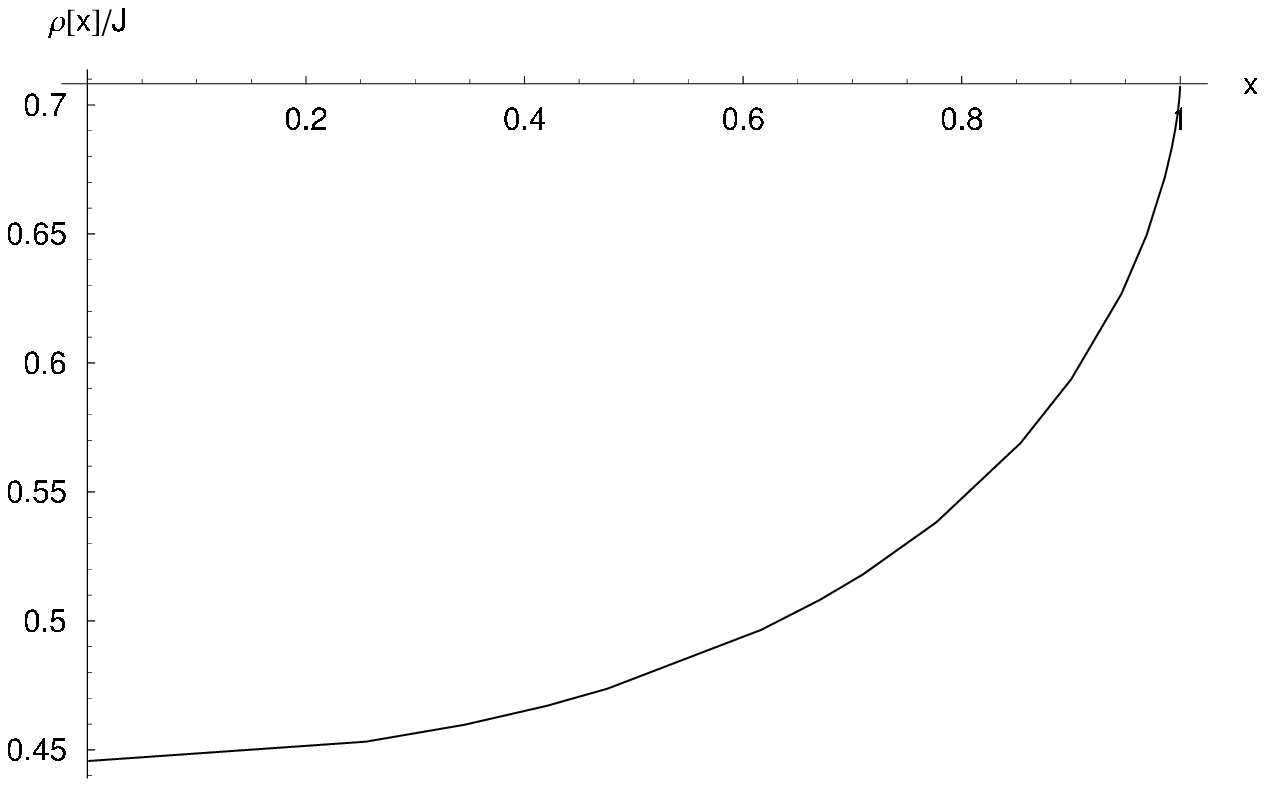} \\ [0.0in] %
\mbox{(a)} & \mbox{(b)} \\ [0.0in]
\end{array}$
\caption{(a) Density of $J$ versus $y=U/U_0$ and (b) Density of $J$
versus $x=X/(L/2)$
 in Eq.~(\protect\ref{eq:Jy}) for $g=0.1$ ($J/J_* \approx 0.9867$.)}%
\label{fig:Jdensity}%
}

In summary, we have found that in the case of infinitely massive
(s)quarks and massless $\Phi$ adjoints, trapping occurs for $J<
J_*\approx 0.22 \sqrt{\lambda}$, decreasing in its effects as $J\to
J_*$.  For $J>J_*$, the adjoints do not bind to the sources, and
the effective potential vanishes.

We have also considered how this calculation is modified if, as in
\cite{maldaloop}, the string's endpoints are at two different
positions on the five-sphere.  Without loss of generality we may
take these positions to lie on the same great circle, one of them at
$+\Delta\chi/2$ and one at $-\Delta\chi/2$. This computation, presented
in Appendix B, shows qualitatively similar behavior; as $\Delta\chi$
increases, the value of $J$ at which the binding energy drops to
zero decreases, until at $\Delta\chi=\pi$, where the state is BPS
saturated, the binding energy is zero even for $J=0$.  The curve in
the $(J,\Delta\chi)$ plane where the binding energy vanishes is
shown in Figure \ref{fig:transition-curve}.  Interestingly, the
$U(X)$ curve and the binding energy are functions of only one
combination of $J$ and $\Delta\chi$, given by the parameter $g$; in
particular, for every $J<J_*$ with $\Delta\chi=0$, there exists a
value of $\Delta\chi$ with $J=0$ that leads to the same curve $U(X)$
and the same binding energy. We do not know of a deep reason for
this.

\subsection{Discussion}

Let us now consider what the Wilson loop computation implies
for the generalized quarkonium states, for which
$m_\Phi$ is not necessarily zero and $m\equiv m_Q$
is finite but much larger than $m_\Phi$.  It is useful to break the discussion
up using a couple of intermediate steps.

For infinite $m$ but nonzero $m_\Phi$, we
expect the new mass-scale to introduce some new physics at a radial
position $U\propto m_\Phi$ in the AdS space.  Since the theory is no
longer conformal, the potential will now be $V(L) =
f(\lambda,J;m_\Phi L)/L$. For instance, if all six scalars receive
positive masses, as in the \nonestar\ theory \cite{nonestar}, then
we expect the AdS space is effectively cut off at $U_{min}\sim
m_\Phi$.  This means that our solutions will be valid only until
$U_0$ reaches $m_\Phi$. This in turn means that the shape of the
string, and the corresponding effective potential, will change once $L>
m_\Phi^{-1}$.  Indeed, the calculation should eventually match on to
that of \cite{annulon}.  For large $L$, the string will lie at
$U=U_{min}$ for most of the region $-L/2 \ <\ X\ <\ L/2$, and we
expect the $\theta$-coordinate of the string to relax to $\pi/2$, as
this should minimize the energy of the configuration. This gives
\cite{annulon}
\begin{equation}
E \approx \sqrt{{\mathcal {T}}^2 L^2 + m_\Phi^2J^2} \qquad
(U_{min}\sim m_\Phi) \ .
\end{equation}
Here ${\mathcal{T}}\sim\frac{\sqrt{\lambda}}{2\pi} m_\Phi^2$ is the
tension of a standard flux tube in the gauge theory.
Thus, the potential
becomes linear in $L$ for sufficiently large $L$ and fixed $J$, and
becomes linear in $J$ for fixed $L$ and sufficiently large $J\gg
\sqrt\lambda$.  The regions with different behaviors are shown in Figure
\ref{fig:crossover}. Note all the transitions are cross-overs; there
are no phase transitions once $m_\Phi$ is nonzero.

\FIGURE[t]{
\epsfxsize=3.7in \epsfbox{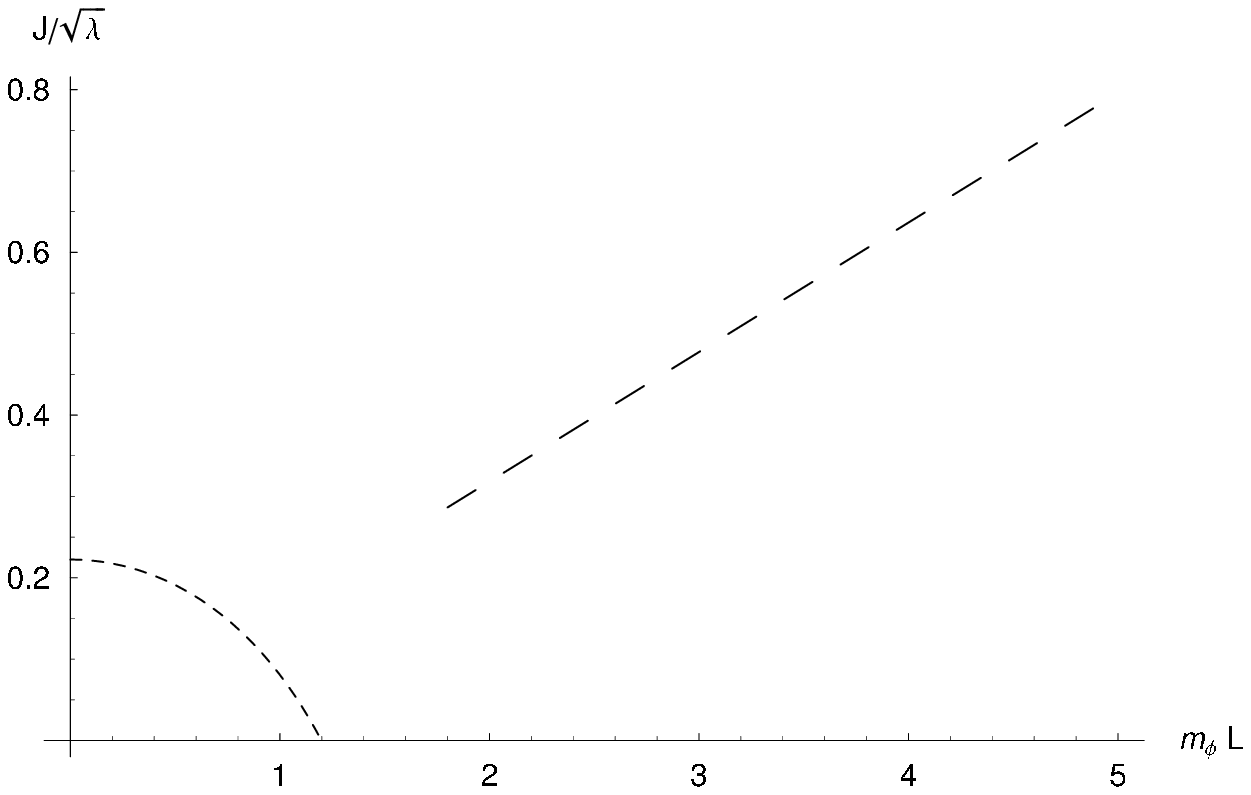}%
\caption{The various regimes for nonzero $m_\Phi$. The diagonal dashed line
is $J = \frac{\sqrt{\lambda}}{2\pi} m_\Phi L$. The potential goes as
$1/L$ in the lower-left region, as $Jm_\Phi$ in the upper left
region, and as $L$ times the flux-tube tension (of order
$\sqrt{\lambda} m_\Phi^2$) in the lower-right region. All borders
between
regions are cross-overs; there are no sharp transitions.}%
\label{fig:crossover}%
}

For finite $m_Q$ and $m_\Phi=0$, on the other hand, the static Wilson
loop we have been studying is no longer physical; instead the (s)quark
and anti(s)quark must be put in orbit around each other.  We have not
attempted to examine this more complex dynamical problem carefully for
general $J$, \new{but in the regime $J\approx J_*$ the situation is
under control.  As we have discussed earlier, in this regime the force
between the (s)quark and anti(s)quark is small and their motion is
slow and nonrelativistic, justifying a Born-Oppenheimer approach.  The
effective potential for the slow (s)quark and anti(s)quark is given by
the Wilson loop calculation we just performed.  As long as
$\alpha_{{\rm eff}}\equiv EL$, as given in Eq.~\eref{ELJstar}, is
small compared to 1 (as becomes true in the regime $J\approx J_*$),
then we would expect that the heavy particles will move with a
velocity $v\sim \alpha_{{\rm eff}}\ll1$, as required.\footnote{There
will be a full Coulomb spectrum of states in this regime.  Of these,
only a handful are adiabatically related to the states with adjoint
trapping, in particular only those states with spin $\leq 1$ and
radial excitation $n$ with $J+n < \lambda^{1/2}$ \cite{myers}.  The
rest of the states are never described as supergravity states for any
$\lambda$, requiring instead the full string theory; correspondingly
they are always much larger in size than the trapped states.}  }

\new{
However, to see whether and when this computation is valid, we must
self-consistently solve for $L$, given a fixed quark mass $m_Q$. The
scaling of the various quantities as $g\to 0$ is now different from
the scaling relation \eref{scaling}, which was obtained holding $L$
fixed.  Since $L = (\alpha_{{\rm eff}} m_Q)^{-1}$ in a Coulomb
system and $U_0\sim \sqrt{\lambda}g/L$, the scaling of the various
quantities with $g\sim (J_*-J)^{1/2}\lambda^{-1/4}$ as $g\to 0$ is
\footnote{See Eq.~(\ref{eq:neartransition}) in the Appendix A.}
$$
\alpha_{{\rm eff}}= EL\sim \lambda^{1/2}g^3, \quad L \sim
\lambda^{-1/2}g^{-3}, \quad E\sim \lambda g^6, \quad U_0\sim \lambda
g^4  \ .
$$
If we only required $\alpha_{{\rm eff}}\ll1$, we would allow
$J_*-J\ll \lambda^{1/4}$.  But this is not sufficient.  We must also
require $U_0\ll m_Q$, so that our Wilson line computation with
strings that extend to infinity bears some relation to the
calculation with strings that extend only to the D7 brane at
$U=m_Q$.  Furthermore, we must also require $L\gg
\sqrt{\lambda}/m_Q$, the trapping scale, which we showed in
\cite{hadsize} also sets the scale of the (s)quark and anti(s)quark
wave functions for the trapped states.  The last condition requires
$\alpha_{{\rm eff}} \ll 1/\sqrt{\lambda}\ll1$.  This in turn
requires $J_*-J \sim \lambda^{-1/6}\ll 1$!  This means that for most
values of $\lambda$ there is no integer $J$ for which there is a
state described by a Born-Oppenheimer calculation.  }

\new{ Still, this scaling is still sufficient for our purposes: for
fixed $J$, there is always some range of $\lambda$ for which the
calculation is valid.  If we follow any particular state of definite
$J$ as $\lambda$ is decreased adiabatically, the state becomes unbound
at $J=J_*$ by passing, rather swiftly, through a regime in which it is
described by our Born-Oppenheimer calculation, grows much larger than
its typical size $\sqrt\lambda/m_Q$, and unbinds when $J_*\sim
\sqrt\lambda$ reaches $J$.  Consequently, for any fixed $\lambda$,
there are no bound states with $J>J_*$.
All states with low
spin, no radial excitation, and $J_*-J>\lambda^{-1/6}$
have adjoints trapped in a small region.  This is consistent with
our earlier conjectures; what we have learned with these calculations
is how rapidly the transition occurs.
}

\new{ Finally, let us consider what will happen if both $m_\Phi$ and
$m_Q$ are nonzero, with $m_{\Phi}\ll m_Q$.  In this case we expect a
cross-over, as in the Wilson-loop computation, to occur in the regime
$J\approx J_*$.  As before,
the Born-Oppenheimer regime exists at $J_*-J\sim \lambda^{-1/6}$.
The calculation of the Coulombic potential and its coefficient
need only be modified by confinement effects at an even smaller value
of $J_*-J$, at the point where the untrapping of the adjoints allows
them to reach the confinement scale $\Lambda\sim 1/m_\Phi$.\footnote{More
precisely, for the above statements to be true, we require
$m_\Phi\ll m\lambda^{-5/6}$.  Moreover
when this condition is satisfied, the calculation's validity
extends somewhat further, since the (s)quark wave functions are smaller at this
transition point, their size being set by a geometric combination of
$1/m_\Phi$ and $1/m_Q$.  However the physics of the states beyond the
transition point as we take $J>J_*$ is more complicated, and we will
not discuss them further here.}  Again there is a
cross-over, a rather sharp one, in which, for fixed $J$ and
decreasing $\lambda$, each bound state goes from a object of size
$\sim 1/m_Q$ with trapped adjoints to a larger confined state of a
size $\sim 1/m_\Phi$.  Equivalently, at any fixed $\lambda$, there is
a sharp transition in the spectrum where low-spin
zero-radial-excitation states with $J<J_*$ have the adjoints trapped
in a region of size $\sim 1/m_Q$, while states with $J>J_*$ spread
their adjoints over a region of size of order $1/m_\Phi$, with the
(s)quark wave functions somewhat more compact.  }

Of course, all of these results are modified at finite $N$.  We expect
the potential energy does not change dramatically.  A more important
effect is that the strings of moderate $J$ ($1\ll J\ll\sqrt\lambda$)
and $L\ll 1/m_\Phi$ can decay, by emission of closed strings carrying
nonzero charge $J'$, to open strings with charge $J-J'$.  In other
words, generalized quarkonium states can decay, via emission of states
of the \nfour\ or \nonestar\ theory carrying the global $U(1)$ charge.
The typical closed string carrying charge $J'$ will correspond (in a
confining theory --- recall that the confinement scale $\Lambda\sim
m_\Phi$ for large $\lambda$) to a hadron with mass of order
$J'm_\Phi$.  Meanwhile, for a string with endpoints separated by a
length $L$, $\partial E/\partial J\sim 1/L\gg m_\Phi$, implying that
these decays are kinematically allowed.  The same is true for
dynamical generalized quarkonium, \new{for which $L\sim \sqrt{\lambda}
m_Q^{-1}$ when $J$ is not near $J_*$.}  The phase space for these
decays is substantial, but by dialing the coupling $1/N$ the widths of
the generalized quarkonium states can be made arbitrarily small
compared to their masses.  We would therefore expect them to remain as
sharp resonances for large but finite $N$.

Finally, let us address the issue of how the physics of large
$\lambda$ matches on to that at small $\lambda$. Since trapping occurs
for $J < J_*\sim \sqrt{\lambda}$, it simply need not occur for
theories with $\lambda < 1$.  This agrees with our understanding of
these generalized quarkonium states in perturbation theory, which
suggests that their size should be of order $m_\Phi^{-1}$ and that
they should become unbound as $m_\Phi\to0$.  However, it is worth
considering the possibility that the $J=1$ state in a QCD-like theory
with additional adjoint matter might, in very favorable circumstances,
exhibit adjoint trapping.  The quark masses $m_Q$ would need to lie
not far above $\Lambda$ (so that $\lambda(m_Q)\sim 1$) with the
adjoint masses lighter than $\Lambda$, and perhaps additional binding
interactions (such as a Yukawa interaction between the quarks) might
also be needed.  This possibility could be explored numerically,
though rather large $N$ might be required in order to stabilize the
state against decay.  Although a long-shot, the observation of adjoint
trapping in lattice gauge theory simulations would certainly be
remarkable.

\acknowledgments We thank Andreas Karch, Charles Thorn and Ariel
Zhitnitsky for useful conversations. We also thank A. Tseytlin and
K. Zarembo for comments on the manuscript.  We also thank the referee
for encouraging us to clarify our arguments. This work was
supported by U.S. Department of Energy grants DE-FG02-96ER40956 and
DOE-FG02-95ER40893, and by an award from the Alfred P. Sloan
Foundation.

\appendix

\section{The behavior near $J=J_*$}

To get the behavior of $J$ near $J_*$ as $g \rightarrow 0$, we expand each
side of Eq.~(\ref{eq:theta-c}) with respect to $\epsilon
\equiv \sin\theta_*-\sin\theta_0$ and
$g^2$ respectively

\begin{equation}
(\sin\theta_* - \epsilon)K(\sin\theta_* - \epsilon) = \sqrt{1 -g^2}
\int_{1}^{\infty} \frac{dy}{\sqrt{(y^2-1)(y^2+g^2)}}
\end{equation}
where $\sin\theta_*$ is determined by Eq.~(\ref{eq:theta*}). After
expanding,
\begin{equation}
\epsilon = \frac{3\pi/8}{K(\sin\theta_* ) + \sin\theta_*
K'(\sin\theta_* )} g^2 + O(g^4) \ .
\end{equation}
Expansion of $J$ near $J_*$ gives
\begin{equation}
J_* - J = \epsilon  \frac{\Rs}{\pi}[K'(\sin\theta_*
)-E'(\sin\theta_* )] + O(\epsilon^2) \ .
\end{equation}
Therefore, as $J \rightarrow J_*$ ($g \rightarrow 0$),
\begin{eqnarray} \label{eq:neartransition}
\sin\theta_*- \sin\theta_0 & \approx &
\frac{3\pi/8 }{E(\sin\theta_*)/\cos^2\theta_*} g^2 \nn \\
J_* - J & \approx & \frac{\sqrt{\lambda}}{\pi} \frac{\sin\theta_*
E(\sin\theta_*)}{\cos^2\theta_*}
(\sin\theta_*- \sin\theta_0) \nn\\
EL & \approx & - \frac{\pi\sqrt{\lambda}}{8}g^3 \approx - \pi
\left(\frac{2}{3}\right)^{3/2} \sqrt{\lambda} \left(\frac{J_*
-J}{\sin\theta_* \sqrt{\lambda}}\right)^{3/2} \ .
\end{eqnarray}

It is straightforward to get the behavior as $J \rightarrow 0$ ($g
\rightarrow 1$),
\begin{eqnarray}\label{eq:asymps}
\sin\theta_0 & = & \frac{ [\Gamma(1/4)]^2}{2 \pi^{3/2}} (1-g)^{1/2}
+ O[(1-g)^{3/2}] \nn \\
J & = & \frac{\Rs}{4}\sin^2\theta_0 + O(\sin^4\theta_0) \nn \\
EL & \approx & E_0 L + \zeta J \ .
\end{eqnarray}
where $E_0 L = - 4 \pi ^2 \Rs / [\Gamma(1/4)]^4$ and $\zeta = 8
\pi^3 / [\Gamma(1/4)]^4 $~.

 From the expansion of $I(g,y)$ for small $g$,
\begin{eqnarray}
I(g,y) &\!\!\! = \!\!\!& \half \left( \frac{\sqrt{y^2-1}}{y} +
\sec^{-1}(y) \right) -\frac{g^2}{16 } \left(
\frac{(3y^2+2)\sqrt{y^2-1}}{y^4} +
3 \sec^{-1}(y) \right) + O(g^4) \nn \\
&\rightarrow& \left\{ \begin{array}{ll}
                        \left(1-\frac{g^2}{2}\right) \sqrt{2(y-1)} +
O[g^4,(y-1)^{3/2}] \quad , \quad y \rightarrow 1 \\ \\
                        \frac{\pi}{4}- \frac{1}{3y^3} - \frac{3\pi g^2}{32} +
O(g^4,1/y^5) \quad , \quad y \rightarrow \infty
\end{array} \right.
\end{eqnarray}
where $y=U/U_0$,
we can get the shape of the string near $U=U_0$ or
near the boundary ($U/U_0 \rightarrow \infty$) as $J \rightarrow
J_*$,
\begin{eqnarray}
x^2 & = & \frac{32}{\pi^2} \left(1-\frac{g^2}{4}\right)(y-1)  +
O[g^4,(y-1)^{3/2}] \quad , \quad y \rightarrow 1 \\
1-x & = & \frac{4}{3\pi} \left( 1+ \frac{3g^2}{8} \right)
\frac{1}{y^3} + O(g^4,1/y^5) \quad , \quad y \rightarrow \infty
\end{eqnarray}
where $x=X/(L/2)$.

\section{String Stretched Between Two D7-Branes at Different Angles}
We consider a stationary string configuration rotating on $S^3$ and
with its ends on two parallel D7-branes, located at $U=\infty$
with angular separation
$\Delta \chi$ (here $\chi$ is the polar coordinate
in the $x^8-x^9$ plane, as in the metric
(\ref{metric}).)

After using conformal gauge and fixing $\tau = t$
\begin{equation}
\mathcal{L}= \frac{U^2}{\Rs}(1+x'^2) + \frac{\Rs}{U^2}U'^2 + \Rs
(\theta'^2 + \cos^2\theta\ \chi'^2 - \omega^2 \sin^2\theta ) \ ,
\end{equation}
where $'\equiv \partial_\sigma$.
Again the equations of motion for $U,X$, which are the same
as for $\Delta\chi=0$, decouple
from the equations of motion for $\theta,\chi$:
\begin{eqnarray}
X'&=& g U_0^2/U^2\\ \nonumber \\
\frac{U'^2}{U^2} + \frac{g^2}{\Rq}\frac{U_0^4}{U^2} -
\frac{1}{\Rq}U^2 &=& \frac{g^2-1}{\Rq}U_0^2 \\ \nonumber \\
\chi' &=& l \frac{U_0 \cos\theta_0}{\Rs}\frac{1}{\cos^2\theta}
\label{eom:chi}\\
\nonumber \\
\theta'^2 + \omega^2 \sin^2\theta + l^2 \frac{U_0^2
\cos^2\theta_0}{\Rq}\frac{1}{\cos^2\theta} &=& \omega^2
\sin^2\theta_0 + l^2 \frac{U_0^2}{\Rq} \ ,
\end{eqnarray}
where $g$ and $l$ are (dimensionless) constants of integration.
$\theta_0$ and $U_0$ are the values of $\theta$ and $U$ at
the midpoint of the string, $X=0$, where we take $\chi=0$.
At the endpoints of the string $X\to\pm L/2$,
$\theta\to 0$, $U\to\infty$,
and $\chi\to \pm\Delta\chi/2$.

 The constraint equation resulting from the conformal
freedom of the worldsheet metric in the Polyakov action is
\[
\frac{U^2}{\Rs}(-1+x'^2) + \frac{\Rs}{U^2}U'^2 + \Rs (\theta'^2
+ \cos^2\theta \chi'^2 +
\omega^2 \sin^2\theta
) =0 \ .
\]
This gives the relation
\begin{equation}\label{c}
g^2 + l^2 = 1 - \Omega^2 \sin^2 \theta_0 \quad , \quad \Omega \equiv
\frac{\Rs \omega}{U_0} \ .
\end{equation}
As before, the state has angular momentum $J$ and (divergent)
energy $E$,
\begin{equation}
J = \frac{\delta S_0}{\delta \omega} = \frac{\omega \Rs}{2\pi} \int
d\sigma \sin^2\theta
\end{equation}
\begin{equation}
E = \omega \frac{\delta S_0}{\delta \omega} - S_0 =
\frac{1}{2\pi}\frac{1}{\Rs} \int d\sigma U^2 \ .
\end{equation}

The string configuration is described by the equations\footnote{A
check of our calculation is that the $J=0$ case reduces to Maldacena's
calculation \cite{maldaloop}. From Eq.~(\ref{J}), $J=0$ could
mean two
possibilities: $\theta_0 \rightarrow 0$ or $l/\Omega \rightarrow
\infty$. The former is correct:
$\theta_0 \rightarrow 0$ gives the range
$\Delta\chi = \pi \frac{l/\Omega}{\sqrt{1+(l/\Omega)^2}}$~, while
$l/\Omega \rightarrow \infty$ gives $\Delta\chi = \pi\sec\theta_0$
(valid only at $\theta_0 =0$). Therefore, the $\Omega \rightarrow 0$ limit
is only a measure-zero subset of the $\theta_0 \rightarrow 0$ limit.  We
should take the $\theta_0 \rightarrow 0$ limit first to obtain $J
\rightarrow 0$.

Recalling that $0 \leq \theta \leq \theta_0$, the limit $\theta_0
\rightarrow 0$ ($J\rightarrow0$) makes the integrals on the left
side of Eq.~(\ref{theta-U}) and the right side of Eq.~(\ref{chi-theta})
equal, implying
\[
\frac{\Delta\chi}{2} -\chi = l \int_{U/U_0}^{\infty}
\frac{dy}{\sqrt{(y^2-1)(y^2+1-l^2)}} = l\
F\left(\sin^{-1}(U_0/U),i\sqrt{1-l^2}\right) \ .
\]
We defined the constant $l$ in Eq.~(\ref{eom:chi}) such that it
becomes Maldacena's $l$ \cite{maldaloop} in the limit
$J\rightarrow0$.} :
\begin{equation}\label{theta-U}
\int_0^{\theta} d\theta \frac{\cos\theta}{ \sqrt{(1 -
\frac{\sin^2\theta}{\sin^2\theta_0} )(\cos^2\theta +
\frac{l^2}{\Omega^2 })} } = \sqrt{1-l^2-g^2}
\int_{U/U_0}^{\infty} \frac{dy}{\sqrt{(y^2-1)(y^2+g^2)}}
\end{equation}
which can be written as
\begin{equation}\frac{\Omega}{\sqrt{\Omega^2+l^2}}
\sin\theta_0 \ F\left(\sin^{-1}\left[
\frac{\sin\theta}{\sin\theta_0}\right],
\frac{\Omega\sin\theta_0}{\sqrt{\Omega^2+l^2}}\right) =
\sqrt{1-g^2-l^2}\ F\Big(\sin^{-1}(U_0/U),ig\Big) \ , \nn
\end{equation}
\begin{eqnarray}\label{chi-theta}
\frac{\Delta\chi}{2} -\chi &=& \frac{l\cos\theta_0}{\Omega
\sin\theta_0} \int_0^{\theta} d\theta \frac{1}{\cos\theta \sqrt{(1 -
\frac{\sin^2\theta}{\sin^2\theta_0})(\cos^2\theta +
\frac{l^2}{\Omega^2 })} } \\
&=& \frac{l\cos\theta_0}{\sqrt{\Omega^2+l^2}} \Pi\left(
\sin^{-1}\left[\frac{\sin\theta}{\sin\theta_0}\right],\sin^2\theta_0,
\frac{\Omega\sin\theta_0}{\sqrt{\Omega^2+l^2}}\right) \ , \nn
\end{eqnarray}

\begin{equation}
X = \frac{g \Rs}{U_0} I(g,U/U_0)
\end{equation}
where $I(g,U/U_0)$ is defined in Eq.~(\ref{Ic}).

The angular momentum and the regularized energy are given by
\begin{eqnarray} \label{J}
J & = & \frac{\Rs}{\pi} \int_0^{\theta_0} d\theta \frac{\sin^2\theta
\cos\theta}{ \sqrt{(\sin^2\theta_0 - \sin^2\theta)(\cos^2\theta +
\frac{l^2}{\Omega^2 })} } \\
& = & \frac{\Rs}{\pi} \frac{\Omega^2}{\sqrt{\Omega^2+l^2}}
\left[ K \left(\frac{\Omega\sin\theta_0}{\sqrt{\Omega^2+l^2}} \right)
 -E \left(\frac{\Omega\sin\theta_0}{\sqrt{\Omega^2+l^2}} \right)
\right] \ , \nn
\end{eqnarray}

\begin{equation}
EL= - \frac{2\Rs g^3}{\pi} [I(g)]^2 \ .
\end{equation}

Parameters are determined by the equations matching the boundary
conditions :

\begin{equation}\label{theta0}
\int_0^{\theta_0} d\theta
\frac{\cos\theta}{ \sqrt{(1 - \frac{\sin^2\theta}{\sin^2\theta_0}
)(\cos^2\theta + \frac{l^2}{ \Omega^2 })} } = \sqrt{1-l^2-g^2}
\int_{1}^{\infty} \frac{dy}{\sqrt{(y^2-1)(y^2+g^2)}}
\end{equation}
which can be written as
\begin{equation}
\frac{\Omega\sin\theta_0}{\sqrt{\Omega^2+l^2}} \ K
\left(\frac{\Omega\sin\theta_0}{\sqrt{\Omega^2+l^2}} \right)
=\sqrt{1-g^2-l^2} \ K(ig) \ , \nn
\end{equation}

\begin{eqnarray}\label{chi0}
\frac{\Delta\chi}{2} &=& \frac{l}{\Omega \sin\theta_0}
\int_0^{\theta_0} d\theta \frac{1}{\cos\theta \sqrt{(1 -
\frac{\sin^2\theta}{\sin^2\theta_0})(\cos^2\theta +
\frac{l^2}{\Omega^2 })} } \\
&=& \frac{l\cos\theta_0}{\sqrt{\Omega^2+l^2}} \ \Pi\left(
\sin^2\theta_0, \frac{\Omega\sin\theta_0}{\sqrt{\Omega^2+l^2}}
\right) \nn \ ,
\end{eqnarray}

\begin{equation} \label{U0}
\frac{L}{2} = \frac{g \Rs}{U_0}I(g) \ .
\end{equation}

For given $J$ and $\THETA$, $\theta_0$ and $\frac{l}{\Omega}$ are
determined from Eq~(\ref{J}) and (\ref{chi0}). For these values,
$\Omega$, $l$, and $g$ are determined from Eq~(\ref{theta0}) and
(\ref{c}). For this value of $g$, $U_0$ is determined from
Eq~(\ref{U0}). Finally, $\omega$ is determined from the relation
$\Omega = \frac{\Rs \omega}{U_0}$.

The right hand side of Eq.~(\ref{theta0}) is bounded from above and
the left-hand side can be written as\footnote{This is compatible
with the fact that $J \rightarrow 0$ as $\theta_0 \rightarrow 0$.}
$\frac{\Rs}{\pi}J/\sin\theta_0 + \frac{\Rs}{\pi} \int_0^{\theta_0}
d\theta \cos\theta \sqrt{1 - \sin^2\theta/\sin^2\theta_0
}/\sqrt{\cos^2\theta + l^2/ \Omega^2}$~, where the integral in the
second term is finite. Therefore the value of $J$ is finite for any
value of $g$ and $l$. Since the string configuration in the $X$--$U$ plane and
the corresponding energy have the same expressions as in the
$\Delta\chi=0$ case, $J$ is finite when $U_0$ touches the horizon and
the interaction energy vanishes as $g \rightarrow 0$.
In particular, at $g=0$ the maximum value $\theta_*$ of $\theta_0$
satisfies
\begin{equation}
\frac{\sin\theta_*}{\sqrt{1+l^2\sin^2\theta_*}} \ K
\left(\frac{\sqrt{1-l^2}\ \sin\theta_*}{\sqrt{1+l^2\sin^2\theta_*}}
\right) = \ \frac{\pi}{2} \ , \nn
\end{equation}
where we used $\Omega^2\sin^2\theta_*=1-l^2$ for $g=0$.  This
occurs for finite $J_*$, for any $l$, with $J_*$ reaching zero
when $l=1$ and $\Delta\chi=\pi$.
By numerical
computation, we find the curve in the ($J$, $\Delta\chi$) plane where
$E = 0$; this curve is shown in Figure \ref{fig:transition-curve}.

\FIGURE[t]{
\epsfxsize=3.5in \epsfbox{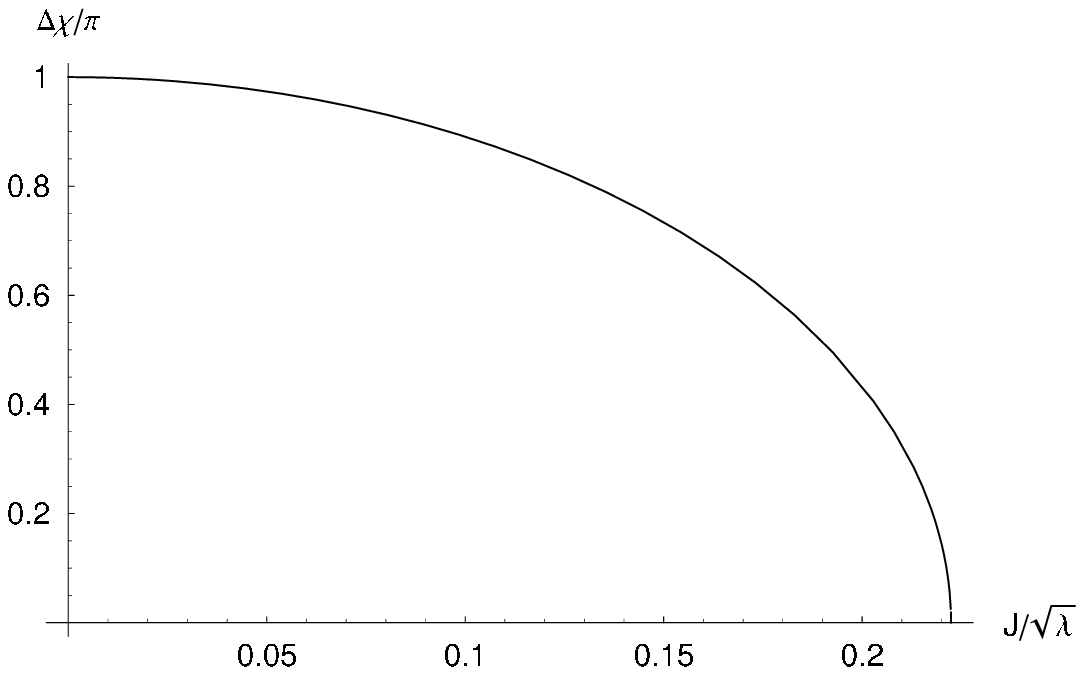}%
\caption{Curve in the ($J$, $\Delta\chi$) plane where $E = 0$
($g=0$).}%
\label{fig:transition-curve}%
}

\section{Elliptic Integrals}
We use the following definitions of the elliptic integrals.
\begin{eqnarray*}
F(\varphi, k) &=&
\int_0^{\varphi}\frac{d\theta}{\sqrt{1-k^2\sin^2\theta}} \\
E(\varphi, k) &=&
 \int_0^{\varphi}d\theta \sqrt{1-k^2\sin^2\theta} \\
\Pi(\varphi, n,  k) &=&
\int_0^{\varphi}\frac{d\theta}{(1-n\sin^2\theta)
\sqrt{1-k^2\sin^2\theta}} \\
K(k)&=& F\left(\frac{\pi}{2},k \right) \\
E(k)&=& E\left(\frac{\pi}{2},k \right) \\
\Pi(n,  k) &=& \Pi\left(\frac{\pi}{2}, n,  k\right)
\end{eqnarray*}
In some references and math-computing programs, $F(\varphi, k^2)$,
$E(\varphi, k^2)$, $\Pi(\varphi, n, k^2)$, $K(k^2)$, $E(k^2)$ and
$\Pi(n, k^2)$ represent the same integral definitions.

\end{document}